# Physical Properties of II Zw 40's Super Star Cluster and Nebula:

# New Insights and Puzzles from UV Spectroscopy

Claus Leitherer

*Space Telescope Science Institute[1], 3700 San Martin Drive, Baltimore, MD 21218, USA*

leitherer@stsci.edu

Nell Byler

*Australian National University, Research School of Astronomy and Astrophysics,*

*Mt. Stromlo Observatory, Cotter Rd, Weston Creek, ACT 2611, Australia*

nell.byler@gmail.com

Janice C. Lee

*Caltech-IPAC, MC 314-6, 1200 E California Blvd, Pasadena, CA 91125, USA*

janice@ipac.caltech.edu

Emily M. Levesque

*University of Washington, C327, Seattle, WA 98195, USA*

emsque@uw.edu

---

[1] Operated by the Association of Universities for Research in Astronomy, Inc., under NASA contract NAS5−26555

# Abstract

We analyze far-ultraviolet spectra and ancillary data of the super star cluster SSC-N and its surrounding H II region in the nearby dwarf galaxy II Zw 40. From the ultraviolet spectrum, we derive a low internal reddening of $E(B - V)$ = 0.07 ± 0.03, a mass of (9.1 ± 1.0) × $10^5$ $M_\odot$, a bolometric luminosity of (1.1 ± 0.1) × $10^9$ $L_\odot$, a number of ionizing photons of (6 ± 2) × $10^{52}$ $s^{-1}$, and an age of (2.8 ± 0.1) Myr. These parameters agree with the values derived from optical and radio data, indicating no significant obscured star formation, absorption of photons by dust, or photon leakage. SSC-N and its nebulosity are an order of magnitude more massive and luminous than 30 Doradus and its ionizing cluster. Photoionization modeling suggests a high ionization parameter and a C/O ratio where C is between primary and secondary. We calculate diagnostic emission-line ratios and compare SSC-N to local star-forming galaxies. The SSC-N nebula does not coincide with the locus defined by local galaxies. Rather, it coincides with the location of "Green Pea" galaxies, objects which are often considered nearby analogs of the galaxies reionizing the universe. Most stellar features are well-reproduced by synthetic spectra. However, the SSC-N cluster has strong, broad, stellar He II 1640 emission that cannot be reproduced, suggesting a deficit of He-enhanced stars with massive winds in the models. We discuss possible sources for the broad He II emission, including very massive stars and/or enhanced mixing processes.

# 1. Introduction

Massive stars are a major powering source for the evolution of stellar systems and galaxies in the local and distant universe. Locally, their detailed physics can be studied in individual objects (Meynet et al. 2017), and in the high-redshift universe their integrated properties serve as tools for exploring galaxy evolution (Finkelstein et al. 2015). Massive stars are accessible over most of the electromagnetic spectrum, either by direct detection or by their interaction with their environment, from gamma-rays (Tanvir 2017), to X-rays (Oskinova et al. 2017), the extreme ultraviolet (UV; Bowyer et al. 2000), the UV/optical/near-infrared (IR; Crowther 2010), mid-IR (Ardila et al. 2010), far-IR (Stock & Barlow 2014), and the radio (Dougherty et al. 2010). Our understanding of massive stars builds on measurements in all wavelength domains. Arguably, the UV region is of particular importance because this is where the spectral energy distribution of an unobscured massive star peaks. Moreover, stellar and nebular physics conspire to provide fewer spectral-line diagnostics outside the UV in hot massive stars: the strongest stellar lines are either exclusively in the UV or they coincide with nebular emission lines in the optical/near-IR (e.g., González Delgado et al. 2002).

The giant H II region 30 Doradus and its central ionizing cluster NGC 2070 in the Large Magellanic Cloud (LMC) have become a benchmark for massive-star studies (Crowther 2017). Key issues which have been addressed include the formation of massive and low-mass stars (Crowther et al. 2010; De Marchi et al. 2017), the stellar initial mass function (IMF; Cignoni et al. 2015), stellar atmospheres and evolution (Vink et al. 2017), and stellar feedback (Pellegrini et al. 2010, 2011). Many more related star clusters have been studied, yet 30 Doradus remains extraordinary as it is the most luminous H II region discovered in the Local Group of Galaxies. Consequently, detecting nearby clusters and H II regions with luminosities and masses similar to, or exceeding those of 30 Doradus and NGC 2070 has become a major quest of massive star research. While any such systems are of interest in their own right, they are invaluable templates for comparison with luminous "super star clusters" (SSC) at larger distances. However, spectroscopically studied SSCs at distances larger than tens of Mpc are typically more luminous than NGC 2070 due to observational biases (e.g., Izotov et al. 2016).

Among the few well studied individual SSCs rivaling NGC 2070 in luminosity are Henize 2-10 (Henry et al. 2007), NGC 3125 (Wofford et al. 2014) and NGC 5253 (Smith et al. 2016). All three have been studied extensively in the UV, both photometrically and spectroscopically, as they are UV-bright. Given the long history of UV observations of massive stars with past and current missions, most high-priority targets have archival UV data available. Recently, Kepley et al. (2014) quantified the star-formation properties of the dominant H II region in the nearby metal- and dust-poor dwarf galaxy II Zw 40 and found it to exceed those of 30 Doradus by a large factor. Panchromatic data for this galaxy are readily available. Surprisingly, UV spectroscopy is lacking. This is mainly because of the high Milky Way foreground reddening, which made UV spectroscopy challenging with previous-generation instruments – but not with the Cosmic Origins Spectrograph (COS) onboard of the Hubble Space Telescope (HST).

In this paper we present COS observations of the giant H II region in II Zw 40 (= UGCA 116) and their modeling and interpretation. Throughout the paper we will refer to the galaxy as II Zw 40, and to the most luminous giant H II region and its most luminous star cluster as "SSC-N", following Kepley et al. (2014). "N" denotes the northern of two massive star clusters. "S", the southern counterpart is 20 times less luminous in Hα (Kepley et al. 2014) and was not observed in our program.

The paper is structured as follows: We begin with a summary of relevant previous work and the derived properties of II Zw 40 in Section 2. In Section 3 we describe the collection of the new COS data and the pipeline processing. The newly obtained data as well as panchromatic ancillary data are presented in Section 4. Section 5 describes the determination of the stellar population using the COS spectrum. The nebular properties are discussed in Section 6. The nature of the broad He II emission in II Zw 40 is the subject of Section 7. The final discussion and the conclusions are in Section 8.

## 2. II Zw 40: Previous Work and Properties

Sargent & Searle (1970) drew attention to the extraordinary nature of II Zw 40 as *dense intergalactic clouds of neutral hydrogen in which the formation of massive stars is proceeding vigorously*, categorizing II Zw 40 as an "extragalactic H II region". A subsequent analysis of the heavy-element abundances in II Zw 40 and the similar I Zw 18 by Searle & Sargent (1972) made these two galaxies the first metal-poor Population I systems to be discovered. Modern determinations of the oxygen abundance in the giant H II region of II Zw 40 lead to log(O/H) + 12 = 8.09 (Guseva et al. 2000), comparable to the value for the Small Magellanic Cloud at about ¼ of the solar value (Asplund et al.

2009). The optical morphology of II Zw 40 shows a core-halo structure with a central dominating H II region surrounded by a halo of gas with two protruding tails as first described by Baldwin et al. (1982). The two tails are reminiscent of other galaxies in various stages of interactions and mergers. The lack of a second galaxy, the kinematic structure and the observed color gradient suggest that II Zw 40 is likely a product of a recent merger and that the current star formation is the direct result of this merger (Brinks & Klein 1988; van Zee et al. 1998).

II Zw 40 has been classified as a blue compact dwarf (BCD) galaxy, a class of galaxies first described by Thuan & Martin (1981). BCDs are characterized by these properties: (i) They have blue colors, although Thuan & Martin use this criterion only qualitatively. For a quantitative definition, see Gil de Paz et al. (2003). (ii) They are compact with optical half-light radii of less than 0.5 kpc. (iii) Their absolute blue magnitudes are fainter than approximately $M_B \approx -18.0$. (iv) Their optical spectra exhibit strong, narrow emission lines powered by ionizing stars. A more recent selection of BCDs by Gil de Paz et al. is based on very similar criteria. As a result of this definition, BCDs are gas-rich, have relatively low stellar masses and, by virtue of the mass-metallicity relation, are metal-poor (e.g., Lee et al. 2004).

Based on these criteria, II Zw 40 is a prototypical BCD. The galaxy is located at Galactic latitude $b = -10.8°$ resulting in a foreground reddening of $E(B-V)_{MW} = 0.73$ (Schlafly & Finkbeiner 2011). II Zw 40 has $(B-R) = 0.77$ and $M_B = -18.1$ after correcting for foreground extinction (Gil de Paz et al.), which makes it one of the more luminous members of its class. The dynamical mass $M_{dyn}$ derived from H I observations is $6 \times 10^9$ M$_\odot$ (Brinks & Klein 1988). There is no standard-candle based distance published in the literature for II Zw 40, however, NASA/IPAC Extragalactic Database (NED) lists a heliocentric radial velocity of $v_{rad}$ = 789 km s$^{-1}$ and a Hubble flow distance of $D$ = 11.1 Mpc. This distance is derived from $v_{rad}$ and the local velocity field model of Mould et al. (2000) using the terms for the influence of the Virgo Cluster, the Great Attractor, and the Shapley Supercluster. At this distance, 1" corresponds to a physical scale of 54 pc. We summarize the fundamental properties of II Zw 40 in Table I. These properties refer to the galaxy as a whole, and not specifically to SSC-N.

II Zw 40's morphology is dominated by the central SSC-N, which is the powering source of a super nebula observed in the optical, IR and radio (Beck et al. 2002, 2013; Kepley et al. 2014). Super nebulae are the very young and gas-rich nebular counterparts of super star clusters (Turner et al. 2000). They are thought to be associated with the earliest phases in the evolution of super star clusters. Despite being embedded in dense gas and dust, the clusters can be probed in the UV due to the patchiness of the interstellar medium (Tremonti et al. 2001 for NGC 5253; Chandar et al. 2003 for Henize 2-10). Super nebulae have also been detected at radio wavelengths in the BCDs NGC 5253 and Henize 2-10 (Beck 2015).

Vacca & Conti (1992) reported strong, *broad* He II 4686 Å emission in II Zw 40, which they interpreted as indicative of Wolf-Rayet (W-R) stars, the hot, helium-rich descendants of very massive O stars (Crowther 2007). Given the short lifetimes associated with W-R stars, broad He II emission is typically weak in the spectra of local starbursts, as expected for a normal IMF. The presence of strong, broad He II 4686 Å emission in II Zw 40 thus suggests the presence of an unusual number of these W-R stars. These stars are extremely hot, and a large population of W-R stars could dramatically influence II Zw 40's ionizing photon budget.

Nebular emission lines provide independent evidence of unusual excitation conditions in II Zw 40. The Infrared Space Observatory (ISO) and the Spitzer Space Telescope detected strong nebular [O IV] 25.9 μm emission (Lutz et al. 1998; Hao et al. 2009). [O IV] emission paired with the presence of nebular, *narrow* He II 4686 Å suggest a very peculiar stellar population, non-thermal processes such as shocks, or even some contribution from a non-stellar powering source (Schaerer & Stasínska 1999; Izotov et al. 2012), since emission from these high ionization species is not typically associated with star formation powered by normal OB stars. Among all metal-poor BCDs with Spitzer data, II Zw 40 has the most extreme [Ne III] 15.6/[Ne II] 12.8 ratio, a ratio often used as an ionization parameter diagnostic. The hardness of the radiation field in II Zw 40 as indicated by the mid-IR [O IV]/Ne II] ratio is similar to that observed in Seyfert galaxies (Fernández Ontiveros et al. 2016).

II Zw 40 is classified as one of the few starburst galaxies in the 11 Mpc local volume, with an enormous Hα equivalent width (EW) of ~450 Å (Lee et al. 2009). SSC-N has been compared to the SSCs in Henize 2-10 and NGC 5253 (Beck et al. 2015) which show similar mass concentrations. The SSCs in these clusters formed at very high efficiencies and will likely evolve into globular clusters. Among the three galaxies, Henize 2-10 has received recent attention for the detection of a candidate black hole (Reines et al. 2011, 2016) or a supernova remnant (Cresci et al. 2017). While there is no evidence for the presence of an active galactic nucleus (AGN) in II Zw 40, the extreme nature of SSC-N makes this galaxy equally extraordinary, and the motivation for our UV study.

## 3. Observations and Data Processing

The COS UV spectra presented here were obtained with HST program 14102 (http://www.stsci.edu/cgi-bin/get-proposal-info?id=14102&observatory=HST) in one visit on January 26, 2016. The archival identifier of the science data is LCXQ01010. The response of the COS far-UV (FUV) detector decreases with usage. In order to mitigate the resulting decreased sensitivity, the spectra recorded by COS are shifted to different detector locations at intervals of several years. The data of program 14102 were taken at lifetime position 3 (LP3; Roman-Duval et al. 2016). COS acquired II Zw 40 through the primary science aperture (PSA) using Mirror A and then centered the PSA on the brightness maximum of SSC-N. Following the target acquisition, four individual spectra were collected over four HST orbits for a total of 11,280 s of exposure time. The four spectra were recorded at different off-set positions shifted by 20 Å in order to calibrate and eliminate fixed-pattern noise. The data were taken in the default TIME-TAG mode at a central wavelength setting of 1105 Å. This setting covers the wavelength range from 1150 to 2200 Å at a nominal resolution of 0.48 Å on Segment A of the COS detector.

We retrieved the complete data set from the Mikulski Archive for Space Telescopes (MAST) and initially processed them on-the-fly with version 3.2.1 of the CalCOS pipeline. The individual CalCOS modules process the data for instrument noise, thermal drifts, geometric distortions, orbital Doppler shifts, count-rate non-linearity, and pixel-to-pixel variations in sensitivity. The wavelength calibration is performed with a standard wavelength scale using the onboard wavelength calibration. The final x1d files generated by CalCOS contain one-dimensional, flux-calibrated, heliocentric-velocity corrected, combined spectra.

We inspected the images taken as part of the acquisition sequence before and after centering. The images are obtained with the near-UV detector over a field of 19.2" × 8.1" in the *x* and *y* direction, respectively. SSC-N was successfully located near the center of the PSA after the initial guide-star acquisition, and its centering was optimized after the target acquisition was complete.

We analyzed the four one-dimensional x1d files for any quality issues. None were identified. The x1d files were then processed and combined with an IDL routine which was developed from a package originally written by the COS Guaranteed Time Observer (GTO) Team (Danforth et al. 2010). This software improves upon the standard CalCOS pipeline by correcting the spectra for major flat-field artifacts in the flux vector, such as ion repeller grid-wire shadows and updates the stored values of the errors and exposure times in the affected pixels accordingly. The decreased signal-to-noise at the detector edges is accounted for by de-weighting these regions. While some of these steps are also done in the default CalCOS pipeline, the GTO software leads to some improvement over the standard data reduction for low signal-to-noise data. We cross-correlated the four x1d spectra and combined them after interpolating the exposure-timed weighted flux vectors onto a common wavelength vector. The spectra were then binned by six pixels from their native dispersion of 0.0803 Å pix$^{-1}$ to 0.482 Å pix$^{-1}$, which equals the nominal point-source resolution. The onboard wavelength calibration was verified by comparing the measured and laboratory wavelengths of the geocoronal Lyman-α 1215.67 Å line. The measured wavelength of 1215.91 Å indicates no significant offset. Additional wavelength zero point shifts can result from the target not being centered and asymmetric target flux profiles in the PSA. These effects can in principle be detected by shifts of Milky Way foreground absorption lines. The Milky Way lines in II Zw 40 are too weak, and the signal-to-noise of the spectrum is too low to perform this test. However, SSC-N is essentially a point source in the UV at the resolution of COS, and there is no indication of a target acquisition failure. We therefore do not expect the actual wavelength scale to significantly deviate from the intrinsic calibration.

## 4. UV and Ancillary Data

In Figure 1 we reproduce archival HST images of II Zw 40 obtained as part of program 9739 (http://www.stsci.edu/cgi-bin/get-proposal-info?id=9739&observatory=HST), retrieved from MAST and processed through the standard pipeline. The images were taken with the Space Telescope Imaging Spectrograph (STIS) using the F25QTZ wide-band filter ($\lambda_0$ = 1596 Å; $\Delta\lambda$ = 232 Å; top panel) and with the Advanced Camera for Surveys (ACS) using the F658N narrow-band filter ($\lambda_0$ = 6584 Å; $\Delta\lambda$ = 78 Å; bottom panel). In both images, SSC-N is centered in the COS PSA (white circle). Note that the bottom panel of Fig.1 is the same image shown in Figure 3 (bottom left) of Kepley et al. (2014).

The UV light of SSC-N is emitted by a cluster of stars, which is fully encompassed by the COS PSA (top panel, Figure 1). No other significant UV light source falls within the PSA. The point source visible outside the southern edge of the PSA is SSC-S, which contributes negligibly to total UV light observed in the PSA.

In contrast to the UV, the F658N image (bottom panel; Figure 1) shows a more extended central region and diffuse emission outside the PSA. The F658N filter includes nebular Hα with some [N II], as well as stellar and some nebular continuum contribution. Despite the more extended Hα emission, we note that the bulk of the Hα emission is still included in the PSA so that a meaningful comparison of the UV and Hα properties can be made.

In Figure 2 we reproduce the processed COS spectrum of SSC-N between 1150 and 2000 Å. No corrections for foreground reddening or redshift have been applied. The drop of the flux towards shorter wavelengths is primarily due to the high foreground reddening of $E(B-V)_{MW}$ = 0.73. This, paired with the Galactic reddening law by Mathis (1990), predicts $A_{1500}$ = 8.2 × $E(B-V)_{MW}$ = 5.99. The declining sensitivity of the G140L grating longward of 1900 Å results in low signal-to-noise at the longest wavelengths. The two strongest emission lines are geocoronal Lyman-α and O I 1304. The measured full-width-at-half-maximum (FWHM) is 8.7 Å, which is expected for spectral lines completely filling the COS PSA. Any Lyman-α intrinsic to SSC-N would be hidden by the geocoronal line.

The strongest stellar (as opposed to interstellar or nebular) spectral lines are N V 1240 and C IV 1550, both of which show P Cygni absorption and emission. These lines are the telltale signs of OB-star winds and are used to study the properties of the recently formed stars (Leitherer et al. 2010). He II 1640 is the only other stellar line which is clearly detected. The broadness of the He II feature clearly implies a non-nebular origin, although some non-stellar contribution may be present. Unlike the N V 1240 and C IV 1550 features, broad He II 1640 emission requires a population of hot stars with massive, fast, He-enriched winds and is generally associated with W-R stars. The presence of He II 1640 is consistent with the detection of broad He II 4686 in the optical by Vacca & Conti (1992). We note that star-forming galaxies often display stellar Si IV 1400 emission or absorption, but this line is not detected in SSC-N.

The most conspicuous nebular emission line is C III] 1908, which has been observed as a strong line in mostly metal-poor star-forming galaxies at low and high redshift (Rigby et al. 2015). Our data do not fully resolve this line into its two components at 1906.7 and 1908.7 Å. We measured an equivalent width of ($-9.5 \pm 0.9$) Å and a line intensity of ($7.6 \pm 0.6$) $\times 10^{-13}$ erg s$^{-1}$ cm$^{-2}$ (corrected for both foreground and internal reddening). This equivalent width is quite high compared to local BCDs (typical CIII] 1906,9 EWs are between 0 and $-15$ Å; Senchyna et al. 2017), and comparable to observations at high redshift (~15 Å; e.g., Erb et al. 2010; Stark et al. 2014). The only other clearly detected nebular emission lines are O III] 1661,66, for which we measured an equivalent width of ($-3.9 \pm 0.4$) Å and a reddening-free line intensity of ($3.5 \pm 0.4$) $\times 10^{-13}$ erg s$^{-1}$ cm$^{-2}$. The values refer to both components and combined and are not corrected for any underlying Al II 1670 interstellar absorption. We also detect

Si III] 1883, but at lower significance. The measured equivalent width and line intensity are ($-1.9 \pm 0.6$) Å and ($1.5 \pm 0.5$) $\times 10^{-13}$ erg s$^{-1}$ cm$^{-2}$, respectively.

The spectrum in Figure 2 also reveals the presence of several interstellar absorption lines, both foreground and intrinsic. The clearest detections are Si II 1260, C II 1335, and Si II 1526. However, the quality of the spectrum is too low for a quantitative analysis of these and other interstellar absorption lines.

We complement the UV spectroscopic data with literature values from optical spectroscopy. Guseva et al. (2000) obtained optical spectrophotometry of II Zw 40 with the GoldCam spectrograph at the Kitt Peak National Observatory 2.1m telescope as part of a larger survey targeting nearby W-R galaxies. Their data were taken with a 2″ slit centered on the brightest part of the galaxy, corresponding in II ZW 40's case to SSC-N and are spatially well-matched to our UV observations given the COS aperture diameter of 2.5″. We adopt the measured line fluxes in their Table 3. Among other lines, detections of the [O II] 3727, H$\delta$, H$\gamma$, [O III] 4363, He II 4686 (both stellar and nebular), H$\beta$, [O III] 5007, [O I] 6300, H$\alpha$, [N II] 6584, and [S II] 6716,31 emission features are reported by Guseva et al. The auroral emission line of [O III] 4363 was used to determine the electron temperature and a total oxygen abundance of log (O/H) + 12 = 8.09.

Kepley et al. (2014) discuss high resolution radio continuum observations with the Very Large Array (VLA) at 4.9 GHz, 8.4 GHz and 22.5 GHz. The super nebula surrounding SSC-N emits weak non-thermal synchrotron radiation compared to the thermal free-free emission in these frequency bands, which is indicative of a low core-collapse supernova (SN) rate. Using Condon's (1992) relation between the synchrotron emission and the *Galactic* SN rate, Kepley et al. derived a SN rate of $4 \times 10^{-4}$ yr$^{-1}$. The 22.5 GHz flux of the super nebula is purely thermal free-free emission. It translates into an ionizing luminosity of $4.67 \times 10^{52}$ photons s$^{-1}$ using the relation of Hunt et al. (2004).

## 5. The Stellar Content of the Central Star Cluster

We determine the stellar properties of SSC-N from a comparison of the UV spectrum with synthetic models generated by Starburst99 (Leitherer et al. 1999; 2014). Prior to the comparison we corrected the data for Galactic foreground reddening $E(B-V)_{MW}$ = 0.73 and applied a redshift correction of $v_{rad}$ = 789 km s$^{-1}$ (see Table 1). The reddening correction was performed with the reddening law of Mathis (1990).

We generated a series of synthetic UV spectra using these inputs and models: We adopted stellar evolution models for massive stars with rotation (Ekström et al. 2012; Georgy et al. 2013). These models are available for two values of the heavy-element abundances, $Z$ = 0.014 and 0.002. The corresponding oxygen abundances are solar and 1/7$^{th}$ solar, respectively. The evolutionary tracks were combined with a theoretical spectral library for OB stars calculated with the WM-Basic code, which accounts for stellar winds in spherically extended, expanding, non-LTE atmospheres (Pauldrach et al. 1998; Leitherer et al. 2010). W-R stars were modeled with the PoWR atmospheres (Gräfener et al. 2002; Hamann & Gräfener 2003, 2004). Stars form instantaneously as a singular event following a Kroupa (2008) IMF, with exponents 1.3 and 2.3 in the mass intervals 0.1 – 0.5 M$_\odot$ and 0.5 – 100 M$_\odot$, respectively. This IMF is essentially identical to a standard Salpeter IMF at the high-mass end, which determines the properties of the UV spectrum.

The comparison between data and models has three adjustable parameters: the intrinsic dust reddening $E(B-V)_{int}$, the total stellar mass $M$, and the age $T$ of the cluster. $E(B-V)_{int}$ is constrained by the continuum slope, which otherwise shows little variation during a phase dominated by O-stars. The additional assumption made is the validity of the adopted dust attenuation law, which in this case is based on Calzetti (2001). The cluster mass is obtained from the reddening-free continuum luminosity. The age is constrained by the time-dependence of the strength of the stellar lines in the synthetic models. We perform least-square fits of the model spectra to the data to determine best fitting values of $E(B-V)_{int}$, $M$, and $T$. There are no evolution models available which exactly match the oxygen abundance of SSC-N. The model with $Z = 0.002$, corresponding to log O/H + 12 = 7.9, is a closer match to the chemical composition of the data and was adopted here.

In Figure 3 we show a comparison of the preferred model (blue spectrum) and the data. The observed spectrum has been corrected for the adopted foreground reddening and the derived intrinsic reddening of $E(B-V)_{int} = 0.07 \pm 0.03$. SSC-N has very little intrinsic dust attenuation; the total observed reddening is almost exclusively due to the high Milky Way foreground reddening.

We derive a cluster mass of $(9.1 \pm 1.0) \times 10^5$ $M_\odot$ for SSC-N, which is an order of magnitude larger than the mass of the star cluster powering 30 Doradus (Doran et al. 2013) and is comparable to the mass of the optically obscured most massive SSC in NGC 5253 (Turner et al. 2017). It is also among the highest values ever observed for young SSCs in the local universe (e.g., de Grijs & Lépine 2010) including those found in the proto-typical Antennae galaxies (Whitmore et al. 2010). The mass is comparable to the masses of the most massive globular clusters in our Galaxy (Gnedin & Ostriker 1997). SSC-N's large mass is still consistent with cluster mass predictions, however. Keto et al. (2005) determined the masses of molecular clouds in the prototypical starburst galaxy M82, and found maximum masses comparable to that of SSC-N.

We derive an age of $T$ = (2.8 ± 0.1) Myr from the fit to the N V, Si IV and C IV lines, marked in Figure 3. N V and C IV weaken with age, as the most massive stars explode as SNe. On the other hand, Si IV is absent early in the cluster evolution but appears after about 3 Myr when the most massive stars leave the main-sequence in the evolution models. The increase in luminosity from dwarf to supergiant stars results in higher mass-loss rates and wind densities, which in turn cause recombination from the dominant $Si^{4+}$ to the trace ion $Si^{3+}$ (Drew 1990). This mechanism is a powerful age discriminant, and the absence of Si IV in SSC-N provides a strong upper limit to the age. The model fit to the N V, Si IV and C IV lines is very good (see Figure 3). Note that changes to the IMF would result in changes to the theoretical profiles. However, no such adjustment was deemed necessary; the standard IMF provides an excellent match to the observations. One should be aware that the age of SSC-N is comparable to the evolutionary time scale of stars more massive than about 150 $M_\odot$ (~2.8 Myr; Yusof et al. 2013). Therefore the data do not constrain stars with masses higher than this value.

Also plotted in Figure 3 is the best-fit model with solar chemical composition (red spectrum). This spectrum is a slightly worse, but still acceptable fit to the data. The corresponding parameters are $E(B-V)_{int}$ = 0.06 ± 0.03, $M$ = (8.1 ± 1.0) × $10^5$ M$_\odot$, and $T$ = (3.0 ± 0.2 ) Myr, which are not significantly different from those derived via the metal-poor model. In the following we will adopt the solution determined from the metal-poor fit.

All other emission and absorption features seen in Figure 3 are not stellar and therefore not included in the synthetic spectra – with the notable exception of He II 1640. The line is commonly ascribed to W-R stars, which are accounted for in both the evolution models and the spectral library. Yet our best-fit model does not predict significant He II at the age of 2.8 Myr because W-R stars have not yet formed. We will return to this failure in Section 7.

The derived age of 2.8 Myr is consistent with other independent age indicators. Kepley et al. (2014) use the relative strengths of the free-free and non-thermal radio emission at cm wavelengths to determine an upper limit of the age of less than ~3.5 Myr. At older ages, core-collapse SNe would lead to an increase of the predicted non-thermal emission. Kepley et al. quote an uncertainty of a factor of a few in their derived supernova rate of 4 × $10^{-4}$ yr$^{-1}$. As a result, their upper limit on the age is not a strong constraint since the models for the SSC-N cluster predict an only slightly higher rate of ~ 5 × $10^{-4}$ yr$^{-1}$ at ages above 4 Myr. However, a young age and an absence of significant numbers of core-collapse SNe is also suggested by the weakness of [Fe II] 1.3 and 1.6 μm in near-IR spectra of II Zw 40 (Izotov & Thuan 2011). These lines are strong in spectra of SN remnants and in starburst galaxies at epochs when core-collapse SNe have exploded. Iron is heavily depleted by dust in H II regions. Shocks associated with SN explosions destroy the dust grains, release Fe, and enhance the [Fe II] lines (Calzetti 1997). The weakness of these lines in SSC-N is again consistent with a young age.

## 6. Properties of the Super Nebula

The Balmer-line fluxes measured by Guseva et al. (2000) permit an independent estimate of the ionizing luminosity of SSC-N. Gusava et al. report an observed Hα flux of $F(H\alpha) = (6.35 \pm 0.08) \times 10^{-13}$ erg s$^{-1}$ cm$^{-2}$ and a Balmer decrement of $F(H\alpha)/F(H\beta) = 6.975 \pm 0.052$. We can use the $F(H\alpha)/F(H\beta)$ ratio for an estimate of the total dust reddening $E(B-V)_{tot} = E(B-V)_{MW} + E(B-V)_{int}$ assuming standard case B recombination for a gas of temperature $10^4$ K and density $10^2$ cm$^{-3}$ (Osterbrock 1989):

$$E(B-V)_{tot} = \frac{2.5}{k(H\beta)-k(H\alpha)} \log \frac{F(H\alpha)/F(H\beta)}{2.86}, \quad (1)$$

where $k(H\alpha)$ and $k(H\beta)$ are the absorption coefficients from the adopted reddening curve. Guseva et al. determined $E(B-V)_{tot} = 0.84$ for a Whitford (1958) extinction law. Using the Cardelli et al. (1989) reddening curve leads to $E(B-V)_{tot} = 0.90$, whereas the Calzetti (2001) attenuation law results in $E(B-V)_{tot} = 0.76$. These values can be compared to $E(B-V)_{tot} = 0.73 + 0.07 = 0.80$, which was found from the slope of the UV continuum in the previous section. The reddening can be used for obtaining dust attenuation corrected values of the observed Hα line fluxes. We find $F_0(H\alpha) = (6.80 \pm 0.09) \times 10^{-12}$ erg s$^{-1}$ cm$^{-2}$, $(5.21 \pm 0.07) \times 10^{-12}$ erg s$^{-1}$ cm$^{-2}$, and $(6.53 \pm 0.08) \times 10^{-12}$ erg s$^{-1}$ cm$^{-2}$ for the dereddened values taken directly from Guseva et al., for correcting them with the law of Cardelli et al., and for correcting them with the Calzetti law, respectively. The mean of the three values with their standard deviation is (6.2 ± 0.7) ×$10^{-12}$ erg s$^{-1}$ cm$^{-2}$. This is the adopted value for the present study. Assuming a distance of 11.1 Mpc this translates into a total Hα luminosity of $L$(Hα) = (9.2 ± 0.1) ×$10^{40}$ erg s$^{-1}$, or a total number of hydrogen ionizing photons of log $N_L$ = $10^{(52.8±0.1)}$ photons s$^{-1}$. $N_L$ derived from the Balmer lines agrees remarkably well with the radio-derived value of 4.67 ×$10^{52}$ photons s$^{-1}$ by Kepley et al. (2014). A third, independent measure of $N_L$ comes from the UV spectral synthesis: by extrapolating the model atmosphere fitted to the observed COS spectrum, one can predict the associated Lyman continuum. The value computed by Starburst99 is $N_L$ = $10^{52.80}$ photons s$^{-1}$, again in excellent agreement with the values determined from recombination data. The nebular, stellar, and dust modeling using radio, optical and UV methods leads to consistent results for the Balmer lines. There is no need to invoke any hidden star formation or non-standard modeling assumptions.

In order to reproduce the non-recombination lines, we computed a series of photoionization models for comparison with the optical and UV emission lines in SSC-N following the method described in Byler et al. (2017, 2018). For stellar evolutionary synthesis, we adopt the Flexible Stellar Population Synthesis package (FSPS; Conroy et al. 2009; Conroy & Gunn 2010) via the Python interface, python-fsps (Foreman-Mackey et al. 2014). We employ the MESA Isochrones & Stellar Tracks (MIST; Dotter 2016; Choi et al. 2016), single star stellar evolutionary models. These models account for stellar rotation. The evolutionary tracks are calculated with the stellar evolution package

Modules for Experiments in Stellar Astrophysics (MESA v7503; Paxton et al. 2011, 2013, 2015). The rotating MESA and Geneva evolutionary tracks differ in that the Geneva models are hotter and more luminous towards the terminal-age main-sequence, suggesting more efficient rotational mixing (Choi et al.). For an age of 3 Myr and less, which is relevant in our case, the predictions for the UV properties of stellar populations made by FSPS and Starburst99 are very similar (Byler et al., Choi et al.). Therefore the stellar synthesis models in the previous section and the photoionization modeling discussed here are fully consistent.

The MIST tracks are coupled with a high-resolution theoretical spectral library (C3K; Conroy, Kurucz, Cargile, Castelli, in preparation) which relies on Kurucz stellar atmosphere and spectral synthesis routines (ATLAS12 and SYNTHE; Kurucz 2005). We extend the C3K library with dedicated libraries for very hot and rapidly evolving stars. For main-sequence stars with temperatures above 25,000 K (O- and early-B-type stars) we use spectra from the BPASS website (Eldridge et al. 2017), which were computed with the WM-Basic code (Pauldrach et al. 1998). For W-R stars, we adopt the spectral library from Smith et al. (2002), which is based on CMFGEN models (Hillier & Miller 1998, 1999).

The photoionization models were computed for a single stellar population with age between 0 and 3 Myr at steps of 1 Myr following a Kroupa IMF. Heavy-element abundances were varied between log $Z/Z_\odot$ = −2 and log $Z/Z_\odot$ = 0.0, and ionization parameters between log $U_0$ = −1 and log $U_0$ = −4, for a fixed inner radius of $10^{19}$ cm (~3 pc) and an electron density $10^2$ cm$^{-3}$. In the following discussion, we compare observed line ratios to the 3 Myr instantaneous burst grid, the closest age match to the best fit age of 2.8 Myr for II Zw 40. We use the intrinsic, unreddened emission-line intensities for SSC-N in all diagnostic diagrams. The optical and UV emission lines of SSC-N were dereddened with $E(B-V)$ = 0.80 and a Calzetti attenuation law.

In Figure 4 we reproduce a "BPT" diagram (Baldwin et al. 1981; Veilleux & Osterbrock 1987) with the computed optical emission-line ratios [O III] 5007/H$\beta$ versus [N II] 6584/H$\alpha$ and the values measured in SSC-N. The figure also shows the location and density of star-forming galaxies from the Sloan Digital Sky Survey (SDSS; Kauffmann et al. 2003). Also included are the data for the sample of metal-poor blue BCDs of Berg et al. (2016), the extreme Green Pea (GP) sample of Jaskot & Oey (2013), the local GP analog Mrk 71 studied by Micheva et al. (2017), and the data for $z \approx 2-3$ star-forming galaxies of Steidel et al. (2014). The SDSS galaxies define the location of galaxies along the sequence from the upper left (metal-poor) to the lower right (metal-rich) as well as the branch to the upper right, where AGN and LINERs are located. II Zw 40 falls well within the star-forming region of the BPT diagram, supporting the idea that the extreme emission observed is not driven by an obscured AGN.

The Berg et al. BCD sample consists of normal metal-poor, star-forming galaxies whose properties are expected to be similar to that of II Zw 40. In particular, these objects have comparable total luminosities, masses and oxygen abundances. Despite these similarities, SSC-N is off-set from the BCD sample towards a larger [O III]/H$\beta$ ratios, suggesting higher excitation conditions. The location of SSC-N coincides with those of the GPs, a group of compact emission-line galaxies at $z \approx 0.1-0.3$ identified in SDSS via their strong [O III] 5007 emission (Cardamone et al. 2009). GPs are considered local analogs of high-redshift galaxies through their low oxygen abundance, low dust content, high UV luminosity, and high specific star-formation rate (Izotov et al. 2011; Amorín et al. 2012). Micheva et al. suggested Mrk 71 as the closest ($D$ = 3.4 Mpc) GP analog. SSC-N and Mrk 71 have nearly identical [O III]/H$\beta$ and [N II]/H$\alpha$ ratios. Like II Zw 40, Mrk 71 is dominated by one powerful star cluster, designated "Knot A" by Micheva et al., which provides most of the ionizing luminosity. Notably, however, the mass and luminosity of Mrk 71's powerful star cluster are an order of magnitude lower than those of SSC-N.

Finally, we emphasize the proximity of the data points of SSC-N and the Steidel et al. (2014) sample in the BPT diagram. These star-forming galaxies at $z \approx 2 - 3$ have "extreme" [O III]/H$\beta$ ratios, offset from the locus of local star-forming galaxies. The cause of the "extreme" [O III]/H$\beta$ ratios in these objects is still debated, but often attributed to an increase in ionization parameter with redshift. Compared with the Steidel et al. sample, SSC-N has a comparable [O III]/H$\beta$ ratio but at much lower [N II]/H$\alpha$ (~0.5 dex). We note, however, that the Steidel et al. galaxies are more metal rich than SSC-N, with chemical compositions ranging between 0.2 < $Z/Z_\odot$ < 1.0. We would thus expect II Zw 40 to have similar excitation conditions to the galaxies in the Steidel et al. sample, but paired with a much lower gas phase metallicity.

In Figure 5 and Figure 6 we show two variants on the standard BPT diagram, using the [O I]/H$\alpha$ and [S II]/H$\alpha$ emission line ratios, respectively. These diagnostics make use of low-ionization lines, and are commonly used to identify LINER objects.

SSC-N does not show excess emission in the low-ionization ratios, and is comfortably within the star-formation region in both Figure 5 and Figure 6. In both diagnostic diagrams, SSC-N is co-located with the GP sample and separated from the sample of metal-poor BCDs.

We detect the nebular lines of C III] 1906,09, O III] 1661,66 and Si III] 1883 in the COS spectrum. We show a UV BPT diagram in Figure 7, using the Si III] 1883/C III] 1906,09 versus O III] 1661,66/C III] 1906,09 ratios, as suggested by Byler et al. (2018). In this figure we compare SSC-N to other galaxies with rest-UV spectra, including the Berg et al. (2016) BCD sample and the sample of low-mass dwarf galaxies at $z \approx 2$ studied by Stark et al. (2014).SSC-N occupies a region that generally coincides with the location of the comparison galaxies, characterized by high values of log $U_0$. Comparison of the location of SSC-N and the model grid lines suggests log $U_0$ = −2.0 ± 0.8 and log O/H + 12 = 7.99 ± 0.20, which compares favorably with our initially adopted oxygen abundance of 12 + log O/H = 8.09. This oxygen abundance is also consistent with the heavy-element abundance of $Z$ = 0.002 adopted for the stellar population synthesis. We note that the theoretical line ratios considered in Figure 7 have low ionization potentials (< 36 eV), and should thus be relatively insensitive to IMF variations. Models with an IMF extending up to 300 $M_\odot$ give virtually indistinguishable results for line ratios involving C III] 1906,09 and O III] 1661,66 in the metallicity regime of interest (Gutkin et al. 2016). This of course no longer applies to line ratios involving high-ionization species such as, e.g., C IV 1550 or N V 1240.

The measured line strengths of C III] 1906,09 and O III] 1661,66 can be used for a determination of the ionic abundance ratio of $C^{++}/O^{++}$. Garnett et al. (1995) and Shapley et al. (2003) derived this ratio from the emission coefficients for collisionally excited emission lines:

$$C^{++}/O^{++} = 0.15 \, \frac{I(\text{C III}]1906,09)}{I(\text{O III}]1661,66)} \, e^{-1.1054/T}, \qquad (2)$$

where $T$ is the electron temperature in $10^4$ K, and $I$(C III] 1906,09 and $I$(O III] 1661,66 are the measured line intensities for the sum of the two components of the C III] and O III] doublets. For $T = 10^4$ K we obtain $C^{++}/O^{++} = 0.108 \pm 0.012$, where the error corresponds to the measurement errors of the two lines. To determine the ratio of the total elemental abundances, we must account for populations in other ionization states using an ionization correction factor (ICF). For SSC-N's estimated $\log U_0 = -2$, photoionization modeling from Erb et al. (2010) indicates that $C^{++}$ and $O^{++}$ are both the dominant ionization states, producing an ICF around unity. As suggested by Erb et al., we adopt ICF = 1.1.

To determine the C/O ratio in the gas and dust phase, we must account for dust depletion. Using the logarithmic depletion factors of −0.30 and −0.07 for C and O from Byler et al. (2018) and and ICF = 1.1 leads to $\log$ C/O = −0.70 ± 0.09. These errors reflect statistical errors and do not account for any modeling uncertainties. The derived C/O ratio is very similar to the values found by Stark et al. (2014) in young low-mass galaxies at $z \approx 2$, as well as in the sample of BCDs of Berg et al. (2016). Shapley et al. (2003) derived $\log$ C/O = −0.68 ± 0.13 in a sample of ~1000 Lyman-break galaxies at $z \approx 3$. Their sample is much more luminous; individual members have luminosities resembling those of *L\** galaxies in the local universe.

We briefly address the presence of *nebular* He II emission in SSC-N. Guseva et al. (2000) detected *nebular* He II 4686 on top of the broad *stellar* He II 4686. The UV spectrum does not have high enough S/N to confirm or exclude the presence of nebular He II 1640 in addition to the broad stellar line. Guseva et al. measured an intensity ratio of 0.016 for He II 4686/H$\beta$, or log (He II 4686/H$\beta$) = -1.8, which is not reproduced in the 3 Myr photoionization model. In fact, at ages less than 3 Myr, the models always have log (He II 4686/H$\beta$) < $-$ 2.5 for any combination of log $U_0$ and log $Z$, indicating that massive stars may not produce enough hard photons. W-R stars have been suggested as the source of hard photons (Schaerer & Stasínska 1999). When the first W-R stars appear around 4 Myr, the photoionization models can produce log (He II 4686/H$\beta$) = $-$1.5, consistent with the nebular emission observed in SSC-N. However, there is no strong correlation between the presence of narrow nebular and broad stellar He II 4686 in W-R galaxies observed in the SDSS (Shirazi & Brinchmann 2012). This casts doubt on the interpretation of nebular He II 4686 (as well as He II 1640) as being powered by W-R stars. Alternative powering sources and emission mechanisms are X-ray binaries, post-asymptotic-giant- branch (AGB) stars or fast radiative shocks but none of these suggestions has been confirmed or rejected.

## 7. Stellar He II 1640 Emission

In this section we will revisit the nature of the He II 1640 observed in the UV spectrum of SSC-N. While broad He II emission in galaxies is commonly interpreted as due to W-R stars (Kunth & Joubert 1985; Conti 1991; Schaerer et al. 1999; Brinchmann et al. 2008), we will consider other mechanisms, such as an active galactic nucleus (AGN), macroturbulence in the interstellar medium (ISM), very massive stars, and binaries as well.

The UV spectrum of SSC-N does not bear any resemblance to Seyfert or quasar spectra, which are characterized by broad emission lines with widths of thousands of km s$^{-1}$ (Kinney et al. 1993). However, the spectrum shown in Figure 2 is strikingly similar to UV spectra of LINERs (Barth et al. 1996, 1997; Maoz et al. 1998), which are thought to be the low-luminosity end of the AGN sequence. He II 1640 in low S/N spectra of NGC 4579 and NGC 6500 is almost indistinguishable from He II 1640 in SSC-N. Could we be observing LINER activity in SSC-N? The answer is not clear on the basis of the UV spectrum alone. However, this interpretation is clearly ruled out by the optical spectrum. By definition, LINERs have enhanced low-ionization lines of, e.g., [O I] 6300 and [S II] 6716, 31 (Ho 2009), whereas in SSC-N these lines have strengths which are fully consistent with an origin in H II regions as discussed in the previous section (see Figure 5 and Figure 6). We conclude that the presence of broad He II in SSC-N does not indicate an AGN.

Alternatively, one might suspect a broadening mechanism in the ISM to be responsible for the width of the He II line. Macroturbulent velocities of hundreds of km s$^{-1}$ have in fact been detected in BCDs (Izotov et al. 2007). However, the observed line profiles are very different, with a narrow core and extended high-velocity wings, which is not seen in He II 1640. More importantly, He II 1640 (as well as its optical counterpart He II 4686) is not the only broad emission feature observed: Guseva et al. (2000) and Buckalew et al. (2005) report the detection of the so-called 4650 feature, a blend of broad C III and N III emission. This C III/N III blend is another telltale of W-R stars, in this case of carbon-rich WC stars. We are therefore left with the conclusion that the He II line in the UV spectrum of SSC-N has a stellar origin.

The comparison between the observed and synthetic spectrum in Figure 3 suggests that the models do an overall reasonable job. The models fail for the W-R line He II 1640 despite W-R stars being accounted for in the models. In order to put the discrepancy on a more quantitative footing, we illustrate the strength and time dependence of He II 1640 predicted by Starburst99 in Figure 8 and Figure 9. The two figures show the behavior of the equivalent width $EW(1640)$ and of the line luminosity $L_{1640}$. The measured values in SSC-N are:

- $EW(1640) = (-7.1 \pm 0.7)$ Å
- $I_{1640} = (7.7 \pm 0.7) \times 10^{-13}$ erg s$^{-1}$ cm$^{-2}$
- $L_{1640} = 10^{(40.06 \pm 0.10)}$ erg s$^{-1}$.

There are four curves in the two figures corresponding to four different sets of stellar evolution models: models with solar chemical composition and 1/7$^{th}$ solar composition ($Z = 0.002$, labeled subsolar in the figures) with no rotation and with a rotation of 40% of the break-up speed on the zero-age main-sequence. The subsolar models with 40% break-up speed were used to derive the properties of SSC-N before, including the spectrum in

Figure 3. The theoretical values of $EW$(1640) and $L_{1640}$ were not derived from an integration of He II 1640 in the synthetic spectrum but in a preferred way following the spectroscopic classification scheme of W-R stars. In short, we classify each W-R star based on the He, C, N and O surface abundance predicted by the evolution models, and then assign a He II line strength as given by Schaerer & Vacca (1998) and scaled by metallicity using the relation of López-Sánchez et al. (2010). For reference, an individual late WN star at solar and $1/7^{th}$ solar chemical composition has $L_{1640} = 1.2 \times 10^{37}$ erg s$^{-1}$ and $7.0 \times 10^{36}$ erg s$^{-1}$, respectively. The data point for SSC-N never reaches any of the models. The most favorable comparison is with the rotating model at solar metallicity, where data and model come closest at ages of 4 and 8 Myr. This model produces the largest number of W-R stars for four reasons: high chemical abundance, high rotation velocity, high luminosity and high mass loss rates, all of which are interrelated (see also Markova et al. 2018). The rise of the curve in this model after 6 Myr is caused by the redward turn of the evolutionary tracks from the red supergiant back to the W-R phase. The gas phase oxygen abundance of SSC-N implies $1/4^{th}$ to $1/5^{th}$ solar metallicity ($Z \approx 0.004$), suggesting that the stellar models at solar metallicity are not the most appropriate comparison. However, the mismatch between the predicted and observed He II strength only worsens when SSC-N is compared to the sub-solar models in Figure 8 and Figure 9 ($Z \approx 0.002$). The subsolar models with and without rotation never produce W-R stars in significant numbers to reach the observed strength of He II in SSC-N. The metal-poor models do, in fact, produce hot, massive stars. However, these stars are not sufficiently enriched in processed material that they would be classified as W-R stars or display W-R features.

While the nebular oxygen abundance for SSC-N implies $Z \approx 0.004$, the $Z \approx 0.002$ stellar models Figures 8 and 9 may actually be the most appropriate comparison. It has been suggested that the stellar and gas phase abundances may be decoupled in objects where a high SFR rapidly enriches the surrounding gas in alpha-elements (e.g., Steidel et al. 2016). Guseva et al. (2000) analyzed the optical nebular spectrum of SSC-N, and obtained an undepleted iron abundance in the gas phase. When combined with the nebular oxygen abundance, this implies an α/Fe ratio of twice solar. Stellar mass loss and evolution of massive stars are driven by the iron, not the oxygen abundance (Vink et al. 2001). Taken at face value, this suggests that the Fe abundance in SSC-N is lower than the gas phase oxygen abundance by a factor of two. Therefore applying models with $Z = 0.002$ (the subsolar models in Figure 8 and Figure 9) would, in fact, be the most appropriate choice.

We have restricted our analysis to single-star evolutionary models thus far. Could evolution in close binaries produce stars emitting broad He II 1640? Eldridge et al. (2017) released the BPASS suite of models, the most comprehensive set of synthesis models of massive stars including binary evolution. Roche-lobe overflow and mass transfer in close binaries leads to harder spectra in massive-star populations compared to populations without binaries. However, this will mainly occur when post-main-sequence stars dominate the population (t > 4 Myr). Eldridge et al. predict the equivalent widths for the optical He II 4686 and C IV 5808 in their models, a signature of the presence of W-R stars. The equivalent widths of the two lines peak at ~4 Å around 3 Myr (C IV 5808) and 3 – 5 Myr (He II 4686). Guseva et al. (2000) measured *EW*(C IV 5808) = 3.5 Å, but only provide an estimate of the total "blue bump" strength (the sum of the C III/N III 4650 feature and He II 4686), with EW = 15.6 Å. Inspection of their plotted spectrum suggests equal strengths of the two components, i.e., *EW*(4686) ≈ 8 Å. This value is about the same as the value for *EW*(1640), which is not surprising as the two lines typically have similar values in W-R galaxies (Chandar et al. 2004). Both the observed C IV 5808 and He II 4686 features are at odds with the predictions of Eldridge et al. (their Figure 28). The theoretical values for *EW*(4686) never exceed 4 Å (a factor of two lower than SSC-N), whereas the theoretical value is *EW*(5808) > 10 Å in the relevant age range (a factor of three larger than SSC-N).

The failure of the synthesis models to reproduce the W-R lines appears to be related to missing ingredients in stellar evolution models. SSC-N is young, therefore almost all stars are still on the main-sequence (as defined by stellar evolution). We suggest that the evolution models used in this study have insufficient mixing processes to produce significant numbers of He II emitters early-on. Such mixing processes can be convection, mass loss or rotation.

How does SSC-N compare to other star-forming galaxies whose broad He II 1640 line has been studied? Chandar et al. (2004) measured the line in a sample of galaxies and SSCs with STIS data. The sample was not selected systematically but is rather composed of galaxies which were bright enough in the UV in order to obtain reasonable

S/N. In Figure 10 we have plotted their measurements. Several galaxies have dual entries, both for the brightest cluster and for the average of all observed clusters. We also added the data point of Crowther et al. (2006) for R136. There is no clear trend with oxygen abundance; $EW$(1640) scatters around an average value of about 3 Å at large and small values of log (O/H). The absence of an oxygen dependence differs from the correlation found for the optical W-R features (Brinchmann et al. 2008) and is most likely the result of biases in the galaxy sample. SSC-N is outside the average for its oxygen abundance but it is clear that several other galaxies or clusters at similar log (O/H) could not be reproduced by the models shown in Figure 8. At oxygen abundances closer to solar, most galaxies would agree with the model predictions if their ages are older than about 4 Myr. Alternatively, binary models are able to match the data for older systems when Roche-lobe overflow has occurred (Eldridge et al 2017). Figure 10 suggests that the modeling challenges we found for He II 1640 in SSC-N are present in other galaxies as well. The discrepancy is more pronounced and harder to resolve in SSC-N with the current models due to its known young age and low oxygen abundance. EW(1640) in SSC-N is comparable to the value for NGC 3125-1, which has the largest value published in the literature. The extreme SSC in NGC 3125 has been studied by Chandar et al., Hadfield & Crowther (2006), and Wofford et al. (2014). This cluster is older than SSC-N, as indicated by the presence of stellar Si IV 1400. Therefore evolution models do predict W-R stars. However, if EW(1640) in NGC 3125-1 is interpreted as due to W-R stars, the required population would rather extreme properties. Wofford et al. favor shortcomings in the models instead. The model failure for He II 1640 in both SSC-N and in NGC 3125-1 suggests that SSC-N is not a peculiar outlier but that our understanding of massive star evolution in general is still incomplete.

The mass range covered by the Starburst99 population synthesis code terminates at 120 $M_\odot$. The models used in this study were restricted to the commonly used IMF upper limit of 100 $M_\odot$, which leads to results indistinguishable from 120 $M_\odot$. If even more massive stars were present following an extrapolated Salpeter-type IMF, their spectral signatures other than He II would be hard to detect in the UV. This would be consistent with the absence of any other spectral anomalies, including the O V 1370 feature, which has been suggested as an indicator of very massive stars (Wofford et al. 2014). Even more massive stars with masses up to 300 $M_\odot$ have been inferred in R136 from a spectral analysis of individual stars (Crowther et al. 2010), which is not feasible at the distance of SSC-N. Smith et al. (2016) suggested the presence of very massive stars in SSC #5 in NGC 5253. Intriguingly, the UV spectrum of this cluster is very similar to that of SSC-N. Its reported *EW*(1640) is (4.2 ± 0.3) Å, and the inferred age is 1 – 2 Myr whereas SSC-N has an age of (2.8 ± 0.1) Myr. Additional support for the age of SSC-N comes from the lack of O V 1371 absorption, which requires the presence of very hot, massive and short-lived stars. In contrast, the line is clearly detected in R136 and NGC 5253 #5. Smith at al. favor an interpretation as due to very massive main-sequence stars with masses of up to 300 $M_\odot$ whose He II is the result of strong mixing and near-homogeneous evolution. This explanation is plausible for R136 and NGC 5253 #5, whose ages are less than 2 Myr. Stellar evolution models predict lifetimes of less than 2.8 Myr for stars with M = 300 $M_\odot$ (Yusof et al. 2013) so that such stars could be present in R136 and NGC 5253 #5, but not in SSC-N.

## 8. Discussion and Conclusions

The new UV spectroscopy obtained with COS permitted a detailed study of SSC-N, the powering source of the BCD II Zw 40. A summary of the derived properties is as follows (cf. Table 2). We find a stellar mass of $(9.1 \pm 1.0) \times 10^5$ $M_\odot$, an age of $(2.8 \pm 0.1)$ Myr, an intrinsic reddening of $0.07 \pm 0.03$, and a bolometric luminosity of $(1.1 \pm 0.1) \times 10^9$ $L_\odot$. These properties follow from the observed absolute flux, the strength of the spectral lines, the slope of the UV continuum and the choice of the stellar evolution models and atmospheres. We adopted single-star models with rotation in the present study. Due to the young age of SSC-N, different stellar evolution models, including those incorporating binaries, make similar predictions for massive stars on the main-sequence. The theoretical spectrum fitted to the UV data can be extrapolated beyond the Lyman break with blanketed, dynamical non-LTE atmospheres to estimate the number of ionizing photons. The predicted value of $N_L = 6.3 \times 10^{52}$ photons s$^{-1}$ is in very good agreement with the numbers derived from recombination data: the consistency with both the radio free-free and optical Hα values indicates that the combined effects of photon escape, dust-hidden star formation and photon absorption by dust are not significant. The young age of SSC-N excludes the presence of large numbers of core-collapse SNe, which is consistent with weak non-thermal radio emission. We derive an upper limit of $< 3 \times 10^{-4}$ yr$^{-1}$ for the SN rate. The UV and optical spectra show broad He II emission.

This emission is generally attributed to He-rich W-R stars with fast, massive winds. The luminosity of the He II 1640 line requires ~1600 mid- to late-WN stars if this line is produced by such stars and if the standard line luminosity calibration is adopted. Stellar evolution models fail to account for such stars.

How do the stellar properties of SSC-N compare with that of the ionizing star cluster of 30 Doradus? SSC-N is barely resolved in HST ACS F814W images. Vanzi et al. (2008) and Hunt & Hirashita (2009) derive cluster radii between ≥ 2.5 pc and ~12 pc. If we assume a radius of ~10 pc, the size of SSC-N is similar to that of R136, the central region of NGC 2070 (Crowther et al. 2016). Despite similar sizes, SSC-N is ten times more massive than R136: Cignoni et al. (2015) determine a total mass of only $9 \times 10^4$ M$_\odot$ for the central 20 pc of NGC 2070 from photometry obtained as part of the Hubble Tarantula Treasury Project. SSC-N's extraordinary properties are also reflected in the feedback of the hot luminous stars on their surrounding ISM. Doran et al. (2013) studied the cumulative radiative and mechanical stellar luminosity in 30 Doradus from a complete census out to a projected radial distance of 100 pc from the center. Doran et al. find a cumulative number of ionizing photons of $N_L = (3 - 8) \times 10^{51}$ photons s$^{-1}$ at a radial distance of 10 pc, depending on the inclusion or exclusion of W-R stars and of photometrically (versus spectroscopically) classified members (their Figure 14). This is again an order of magnitude below the value determined for SSC-N and its surrounding nebula. In conjunction with their ionizing luminosity, hot massive stars shape the ISM with their powerful winds. Starburst99 models predict a wind luminosity of $2 \times 10^{39}$ erg s$^{-1}$ for SSC-N. This can be compared to the cumulative wind luminosity in R136 at a radius of 10 pc. Since our models do not include W-R stars, which have the most powerful winds, our derived value should be compared to the numbers for OB stars in R136 only (the blue-dashed line in Figure 14 of Doran et al.), which is $8 \times 10^{38}$ erg s$^{-1}$. The mechanical luminosity of all OB stars in SSC-N exceeds that in R136 by a factor of 2 – 3. The difference between SSC-N and R136 is less pronounced than in the case of $N_L$ since winds decrease in strength with metallicity, whereas $N_L$ does not.

II Zw 40 is even more extraordinary when placed in the context of the overall population of nearby galaxies in the local volume. Of the 436 galaxies in the 11 Mpc sample of Kennicutt et al. (2008) and Lee et al. (2009), II Zw 40 has the second highest galaxy-wide Hα EW in the sample [(451 ± 23) Å]. The distribution of galaxy Hα EWs exhibits a log-normal distribution, with a mean of 30 Å and a 3 σ width of 100 Å, so II Zw 40 is in the extreme tail of the distribution. Stellar population modeling of the Hα EW shows that II Zw 40 has experienced a starburst which has elevated its star-formation rate to a level which is greater than five times its past average. Moreover, its Hα luminosity, which is mostly powered by SSC-N, is $1.5 \times 10^{41}$ erg s$^{-1}$. This luminosity corresponds to a star formation rate of 1.2 M$_\odot$ yr$^{-1}$, using the basic recipe of Kennicutt (1998). This level of activity is comparable to that exhibited by large disk galaxies such as the Milky Way and M51, except that in II Zw 40, it is packed to an area less than 25 pc across (Vanzi et al. 2008; Hunt & Hirashita 2009), and suffers from very little local dust obscuration. This is the key characteristic of GP galaxies and makes II Zw 40 one of the closest galaxy of this class. While there are a handful of other galaxies in the local volume with comparable Hα EWs, they are far less luminous and have star-formation rates which are a factor of at least 30 times lower (Table 2, Lee et al.).

Engelbracht et al. (2008) derive a total warm dust luminosity of $(1.6 \pm 0.7) \times 10^9$ $L_\odot$ in II Zw 40 from Spitzer Space Telescope photometry. ALMA observations show that the dust is spatially associated with SSC-N (Consiglio et al. 2016). The luminosity of the dust is comparable to the bolometric luminosity derived for SSC-N suggesting the stellar cluster is the powering source. SSC-N would fall short of heating the dust if it were significantly older. For instance, at an age of 20 Myr, corresponding to the epoch when 10 $M_\odot$ stars turn into core-collapse SNe, SSC-N would have $L_{Bol}$ = $8.5 \times 10^7$ $L_\odot$, which is more than an order of magnitude below the dust luminosity. Converting the dust luminosity into a dust mass requires assumptions on properties such as geometry, temperature or grain size and is correspondingly uncertain. Estimates range between $10^4$ to $10^6$ $M_\odot$ (Hunt et al. 2005; Engelbracht et al; Vanzi et al. 2008; Hirashita 2011; Remy-Ruyer et al. 2013; Consiglio et al.; Kepley et al. 2016). What is the source of the dust? The spatial co-location of the dust and SSC-N suggests the producers to be associated with the cluster. Most of the condensed dust observed in the local universe is formed in the winds of AGB stars and in the ejecta of core-collapse SNe (Dwek et al. 2014). The young age of SSC-N excludes AGB stars as the dust source. Similarly, SNe are not numerous enough. Given our derived upper limit, we would expect fewer than $10^3$ SN explosions. Even if each SN event produced 1 $M_\odot$ of dust (Hirashita) and if the dust survived, the total dust mass would fall short by at least an order of magnitude. Hirashita suggested the existence of pre-existing grains in the ISM and subsequent growth of dust via accretion of metals. The time-scale of this formation mechanism is short in comparison with the age of SSC-N. In this scenario, massive O stars would provide the metals via their strong stellar winds as proposed by Consiglio et al. The newly produced dust would be responsible for the anomalously low gas-to-dust ratio. The presence of dust in the vicinity of SSC-N does not completely obscure the starburst at optical and UV wavelengths as we have demonstrated in this study. The reason is likely an inhomogeneous dust morphology, which enables photon escape via a picket-fence structure.

Apart from stellar lines, we detect several key diagnostic nebular lines in the UV spectrum. A standard UV BPT diagram is used in conjunction with new photo-ionization models to determine the nebular properties of the H II region surrounding SSC-N. The gas is characterized by a high ionization parameter of $-2$ and a low carbon-to-oxygen ratio of log C/O = $-0.70$. Carbon and oxygen are produced by different populations of stars on different timescales. While oxygen is produced in Type II SNe from massive stars, carbon is a byproduct of the triple-alpha process, which occurs in both massive and low- to intermediate-mass stars. Thus, the relative abundances of C and O have been studied extensively in galaxies to constrain models of chemical evolution. Recent work from Berg et al. (2016) and Senchyna et al. (2017) found that the C/O ratio was relatively constant with metallicity in low-metallicity dwarf galaxies (12 + log(O/H) < 8), though with large scatter. However, C/O ratios are larger in higher-metallicity star-forming galaxies (e.g., Esteban et al. 2014), suggesting a general trend that C/O increases with increasing oxygen abundance. Because no additional nucleosynthetic carbon production mechanisms exist at high metallicity, the source of the so-called "pseudo-secondary" carbon production remains uncertain. Berg et al. (2016) suggested that higher C/O values could be attributed to enrichment by carbon-rich outflows from low- and intermediate-mass stars, which are strongly dependent on metallicity.

The carbon-to-oxygen ratio of log C/O = −0.70 found in SSC-N is typical for star-forming dwarf galaxies at low metallicity (e.g., Garnett et al. 1995; Berg et al. 2016, Senchyna et al. 2017), which have an average log C/O = −0.6 though with large scatter (± 0.5 dex). For comparison, 30 Doradus, whose oxygen abundance is 12 + log(O/H) = 8.59, has log C/O = −0.45 (Peimbert 2003). SSC-N's C/O ratio is also consistent with C/O values derived for star-forming galaxies at $z$ = 2 – 3 (e.g., Erb et al. 2010; Stark et al. 2014; Steidel et al. 2016). The galaxies at $z$ = 2 – 3 have log C/O between −1.0 and −0.4, with an average value slightly lower than the local dwarf samples (Berg et al. 2018). At 12 + log(O/H) = 8.1, SSC-N is right at the transition between primary and the proposed pseudo-secondary production regimes. The relatively low C/O value for SSC-N paired with the very young age estimate lends support to the idea that carbon enhancement is driven by low- and intermediate-mass stars, which require older stellar populations.

We are confident that the stellar and nebular models are mostly reliable and that the main results found in our study are robust. Nevertheless we need to highlight the puzzling mismatch between data and models for He II 1640, one of the strongest features in the UV spectrum. As we pointed out, this tension also exists for other galaxies (see Figure 10), but it is exacerbated in SSC-N due to its young age and low oxygen abundance. He II 1640 is observed in similar strength in the young star clusters

R136 (Crowther et al. 2016) and NGC 5253 #5 (Smith et al. 2016). In these cases, very massive stars with masses of up to 300 $M_\odot$ have been suggested as the sources. This explanation is less plausible for SSC-N because of its older age when the most massive stars may no longer be present.

Broad He II 1640 emission is indicative of He-rich stars with massive fast winds which may or may not be core-He burning, i.e., they may not necessarily be identified as classical W-R stars. Nevertheless it is instructive to explore which W-R sub-types would be able to account for the line if added to the theoretical spectrum. He II 1640 is present in essentially all WN sub-types observed with IUE and HST (Nussbaumer et al. 1982; Smith & Willis 1983; Bestenlehner et al. 2014; Hainich et al. 2014). However, most WN stars also show very strong N IV 1718, as well as N IV] 1386 and Si IV 1400, which are not seen in the spectrum of SSC-N. These lines are strongest in late-WN stars and tend to be weaker or absent in mid- and early-WN stars. Stars whose spectra resemble WN4-5 stars might mimic the stellar lines if added to the UV spectrum. Their absence in the synthetic spectrum points to a significant shortcoming in the stellar evolution models used here. As these stars may contribute to the ionization and enrichment of SSC-N and other massive star clusters, understanding their nature and their formation channel must be a priority in future stellar modeling.

*Acknowledgments.* Support for this work has been provided by NASA through grant number GO-14102 from the Space Telescope Science Institute, which is operated by AURA, Inc., under NASA contract NAS5-26555. This research has made extensive use of the NASA/IPAC Extragalactic Database (NED) which is operated by the Jet Propulsion Laboratory, California Institute of Technology, under contract with the National Aeronautics and Space Administration.**REFERENCES**

Amorín, R., Pérez-Montero, E., Vílchez, J. M., & Papaderos, P. 2012, ApJ, 749, 185

Ardila, D. R., Van Dyk, S. D., Makowiecki, W., et al. 2010, ApJS, 191, 301

Asplund, M., Grevesse, N., Sauval, A. J., & Scott, P. 2009, ARA&A, 47, 481

Baldwin, J. A., Phillips, M. M., & Terlevich, R. 1981, PASP, 93, 5

Baldwin, J. A., Spinrad, H., & Terlevich, R. 1982, MNRAS, 198, 535

Barth, A. J., Reichert, G. A., Filippenko, A. V., Ho, L. C., Shields, J. C., Mushotzky, R. F., & Puchnarewicz, E. M. 1996, AJ, 112, 1829

Barth, A. J., Reichert, G. A., Ho, L. C., Shields, J. C., Filippenko, A. V., & Puchnarewicz, E. M. 1997, AJ, 114, 2313

Beck, S. 2015, IJMPD, 24, 1530002

Beck, S. C., Turner, J. L., Lacy, J., & Greathouse, T. 2015, ApJ, 814, 16

Beck, S. C., Turner, J. L., Lacy, J., Greathouse, T., & Lahad, O. 2013, ApJ, 767, 53

Beck, S. C., Turner, J. L., Langland-Shula, L. E., et al. 2002, AJ, 124, 2516

Berg, D. A., Erb, D. K., Auger, M. W., Pettini, M., & Brammer G. B. 2018, ApJ, 859, 164

Berg, D. A., Skillman, E. D., Henry, R. B. C., Erb, D. K., & Carigi, L. 2016, ApJ, 827, 12641

**Figures**

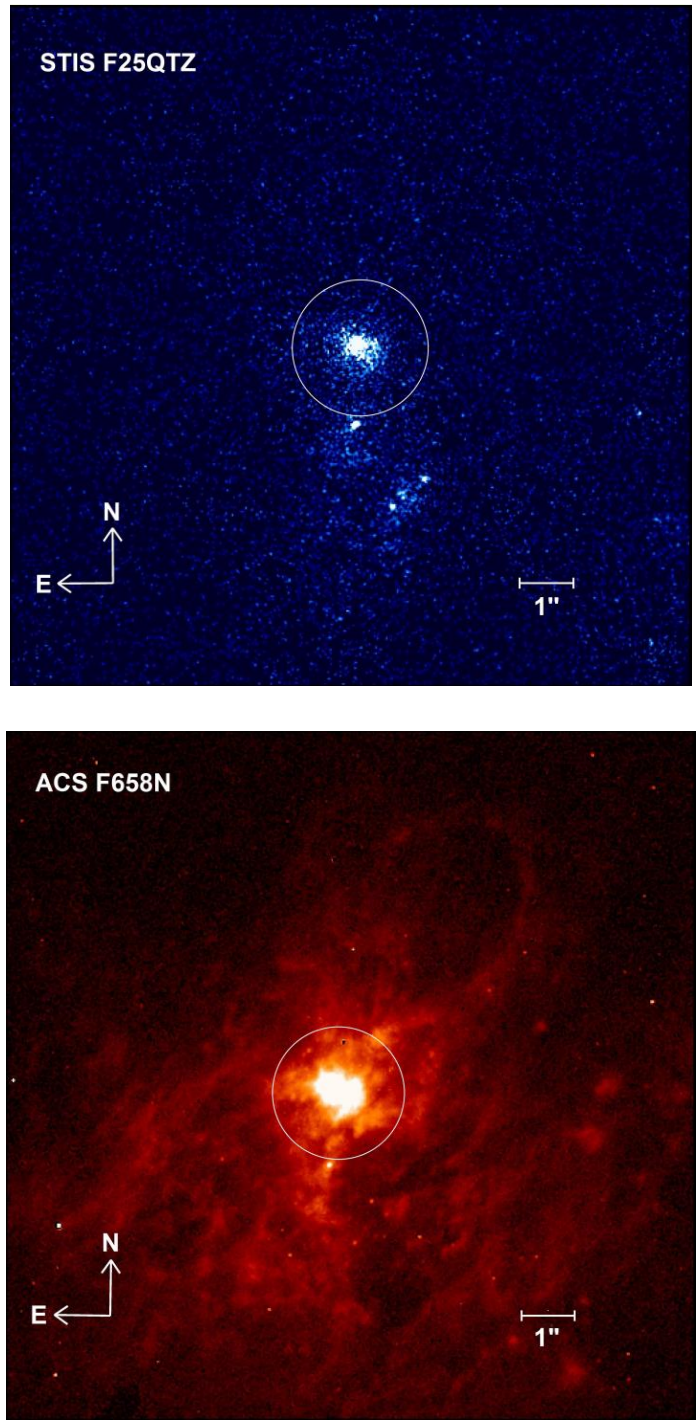

Figure 1. Archival STIS F25QTZ (upper) and ACS F658N (lower) images of SSC-N. The size of the COS PSA is indicated by the circle. The horizontal bar corresponds to 1″.

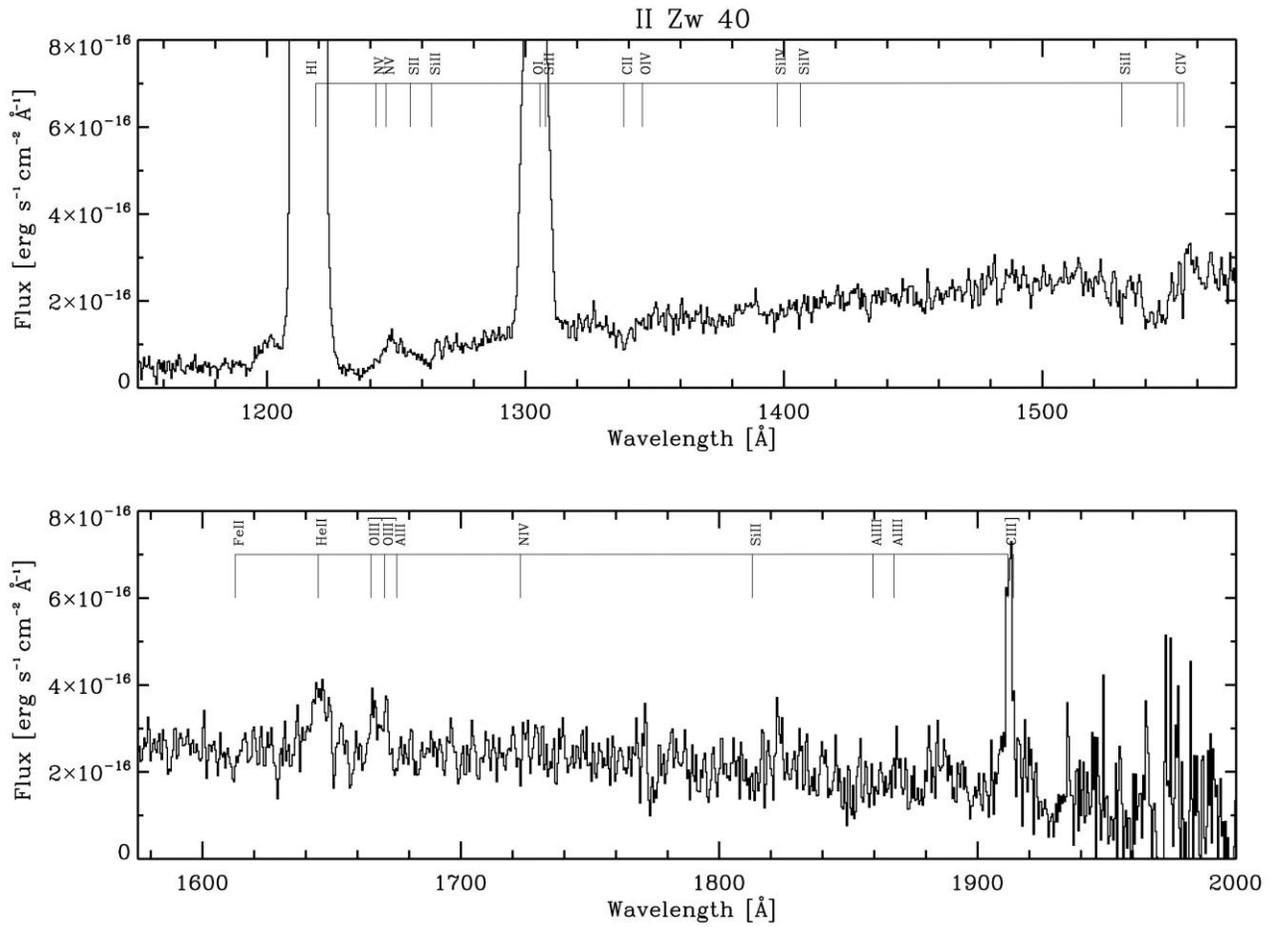

Figure 2. Observed UV spectrum of SSC-N in II Zw 40 between 1150 and 2000 Å. The wavelength scale is in the observed frame; the fluxes are not corrected for reddening. Identifications of spectral features are given at the top of each panel.

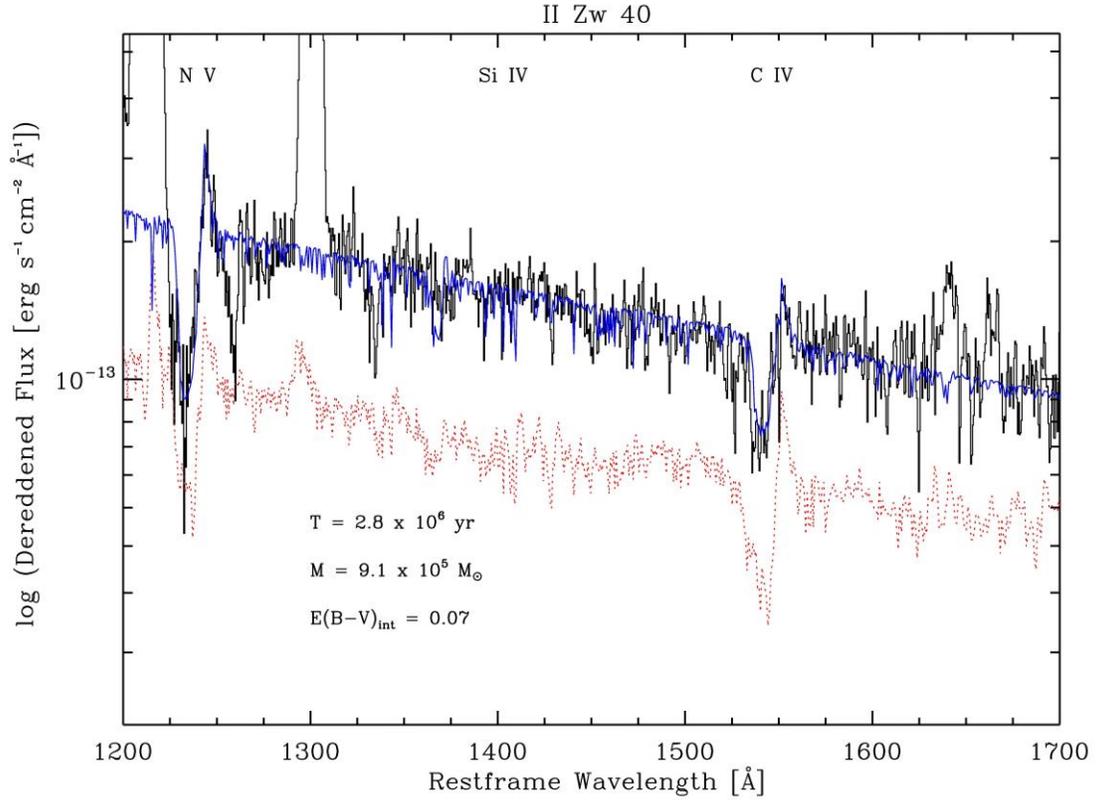

Figure 3. Comparison of the observed spectrum of SSC-N (black) with the best-fit model spectrum (blue). The observed spectrum is in the restframe wavelength and has been corrected for foreground (0.73) and intrinsic (0.07) $E(B-V)$ reddening. The model spectrum corresponds to a star cluster of mass $9.1 \times 10^5$ $M_\odot$, age $2.8 \times 10^6$ yr and chemical composition $Z = 0.002$. For comparison we also show the best-fit model with $Z = 0.014$ (red, dashed). This spectrum has been shifted vertically for clarity reasons.

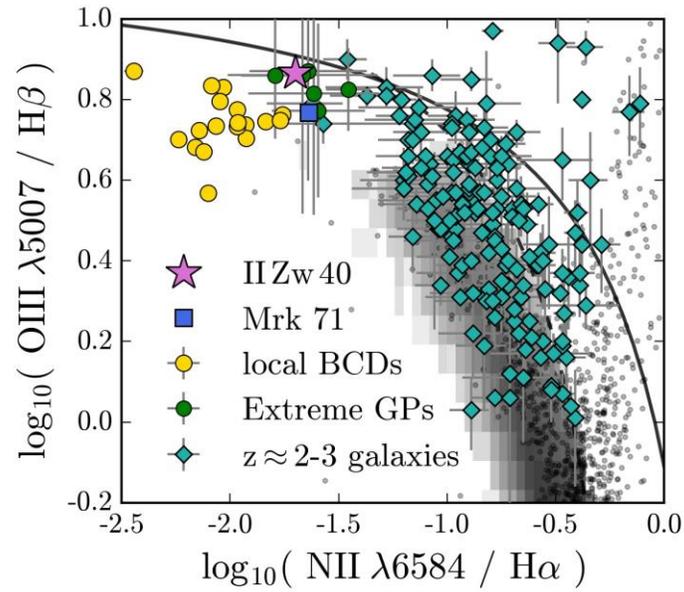

Figure 4. The BPT diagram, using observed emission-line ratios of [O III]/Hβ versus [N II]/Hα. II Zw 40: purple star; BCD sample of Berg et al. (2016): yellow circles; extreme Green Peas of Jaskot & Oey (2013): green circles; Mrk 71 (Micheva et al. 2017): blue square; $z \approx 2 - 3$ galaxies of Steidel et al. (2014): cyan diamonds. The solid black line shows the extreme starburst classification line from Kewley et al. (2001), and the dashed line is the pure star formation classification line from Kauffmann et al. (2003). The grayscale 2D histogram shows the density of SDSS star-forming galaxies, with AGN and LINERs removed using the Kauffmann et al. (2003) classification. The black points show a random sample of SDSS galaxies that have not had AGN or LINER objects removed.

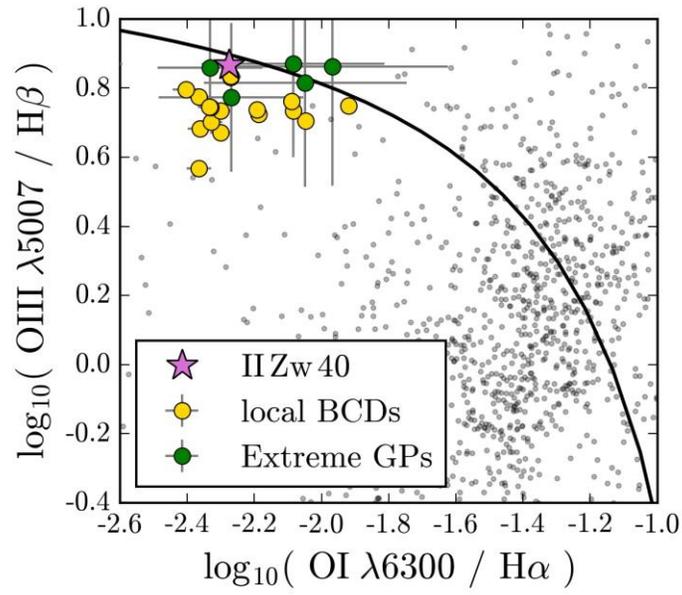

Figure 5. Observed emission-line ratios of [O III]/Hβ versus [O I]/Hα. II Zw 40: purple star; extreme Green Peas of Jaskot & Oey (2013): green circles. The solid black line shows the Kewley et al. (2001) classification for star-forming galaxies. The black points show a random sample of SDSS galaxies.

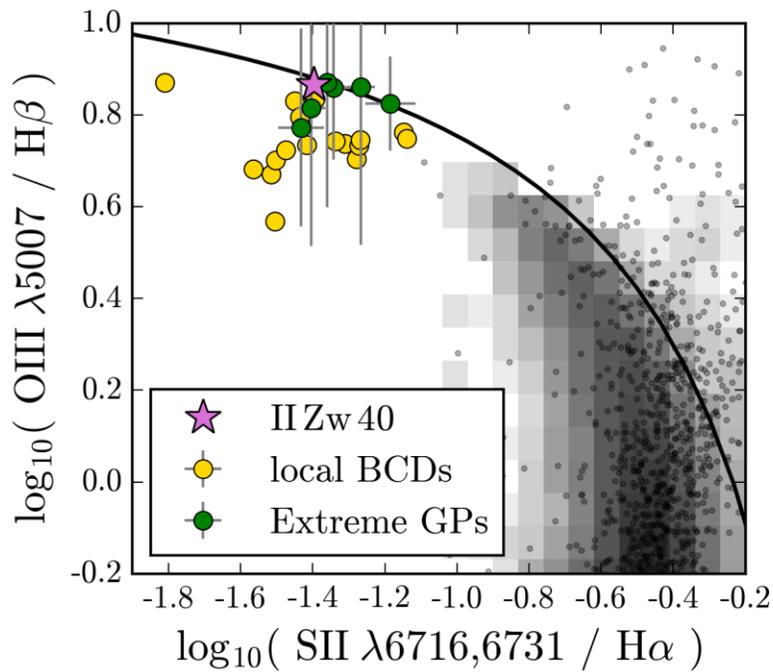

Figure 6. Observed emission-line ratios of [O III]/Hβ versus [S II]/Hα. II Zw 40: purple star; BCD sample of Berg et al. (2016): yellow circles; extreme Green Peas of Jaskot & Oey (2013): green circles. The solid black line shows the Kewley et al. (2001) classification for star-forming galaxies. The grayscale 2D histogram shows the density of SDSS star-forming galaxies, with AGN and LINERs removed using the standard BPT diagram. The black points show a selection of SDSS galaxies that have not had AGN or LINER objects removed.

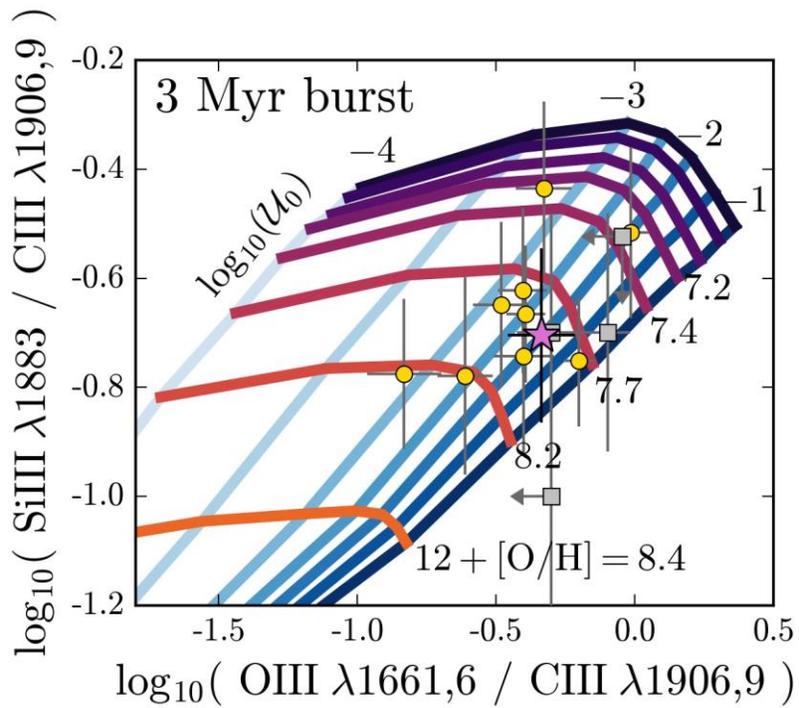

Figure 7. Model emission-line ratios of Si III]/C III] versus O III]/C III], using the emission line model from Byler et al. (2018). The blue lines connect models of constant ionization parameter, from log $U_0$ = −1 (dark blue) to log $U_0$ = −4 (light blue). Models of constant oxygen abundance are connected with colored lines, from dark purple to orange. II Zw 40: purple star; BCD sample of Berg et al. (2016): yellow circles; $z \approx 2$ galaxies of Stark et al. (2014): grey squares.

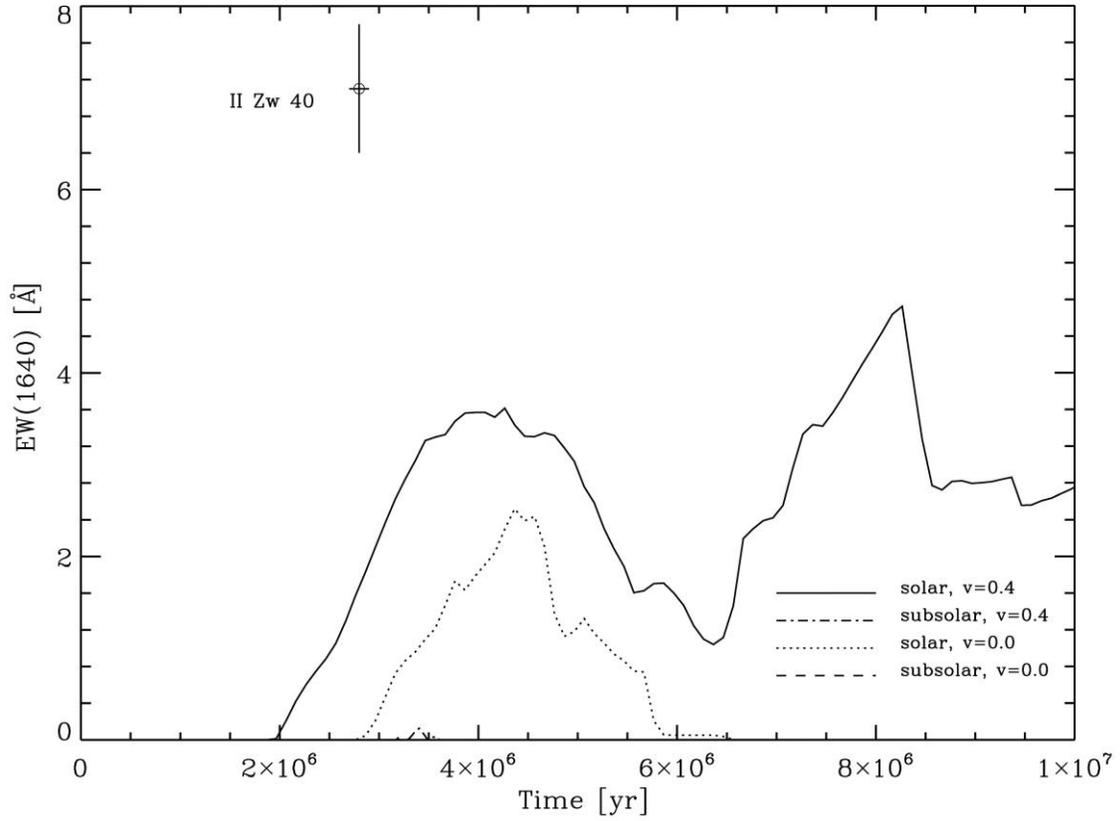

Figure 8. Equivalent width of stellar He II 1640 versus time. The measurement for SSC-N in II Zw 40 is the circle with error bars. The lines show the predicted values from four sets of evolution models: solar and subsolar abundances with and without rotation. The predicted values for the two subsolar models are close to zero at all times and therefore not discernible in the figure.

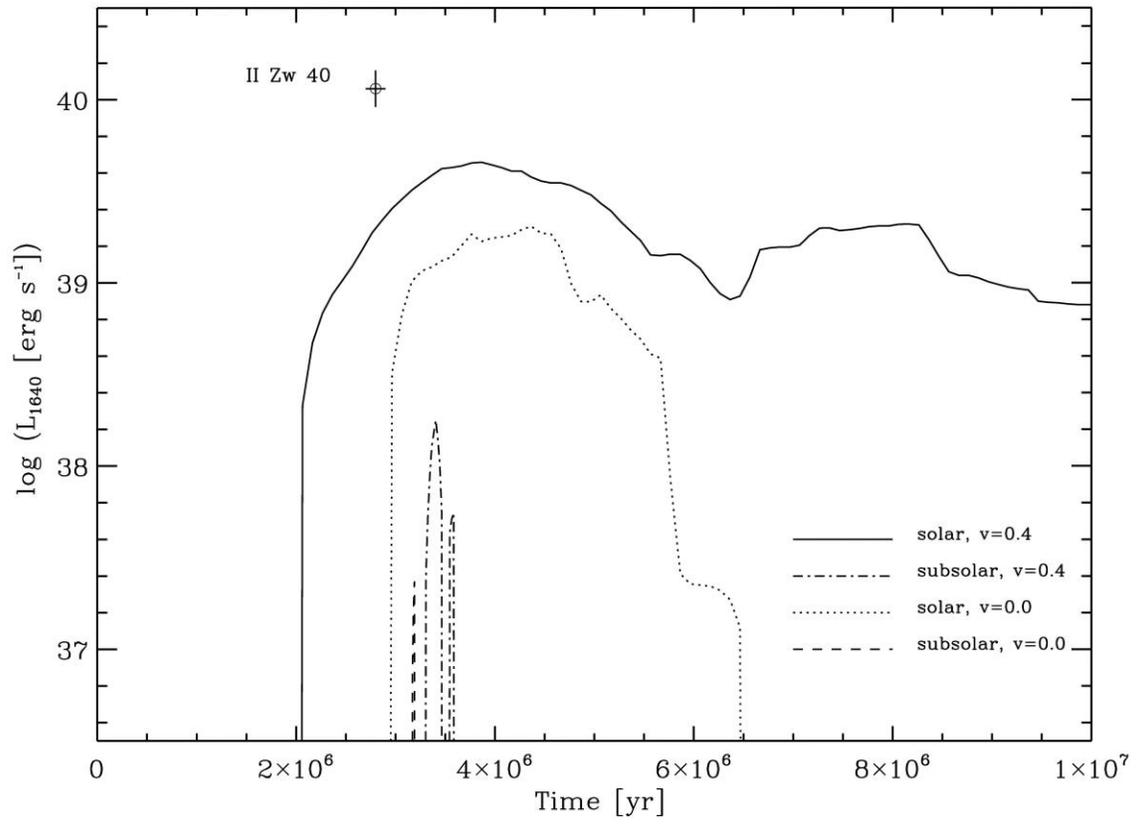

Figure 9. Same as Figure 8 but for the luminosity of He II 1640.

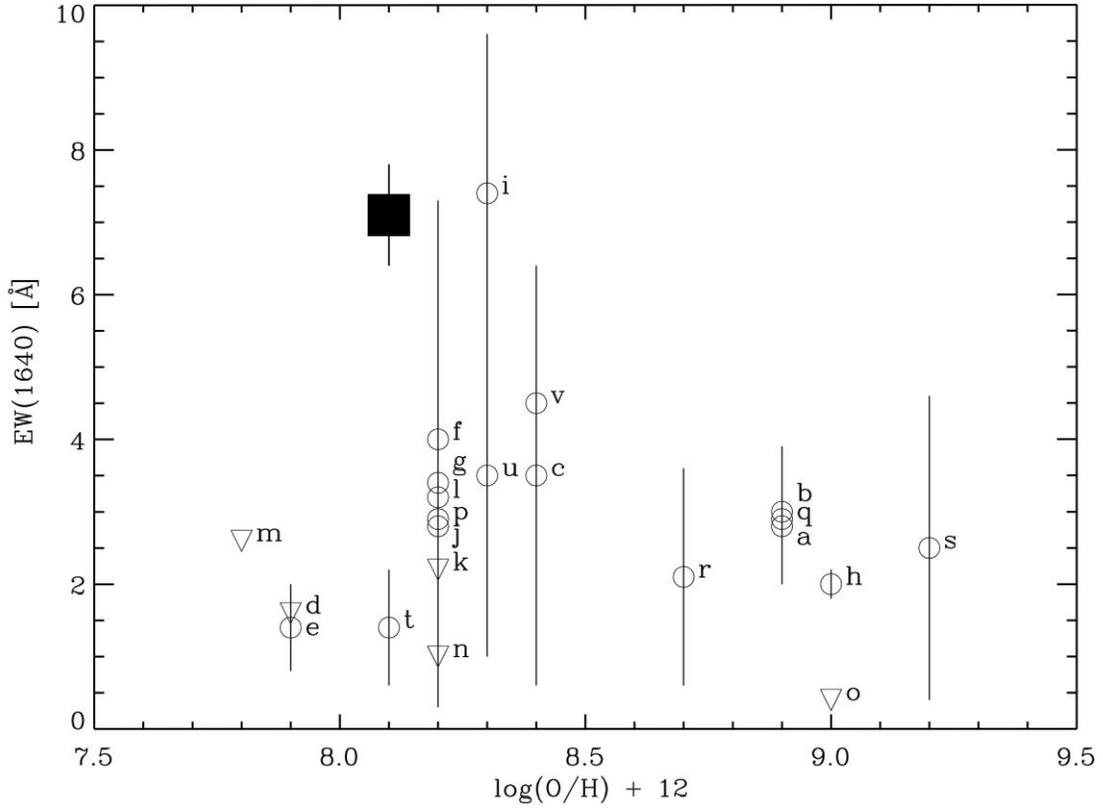

Figure 10. EW(1640) in SSC-N (filled square) compared to literature data. Open circles with error bars are detections; open triangles are upper limits. a – He 2-10-1; b – He 2-10-all; c – Mrk 33-all; d – Mrk 36-1; e – Mrk 209-1; f – NGC 1741-1; g – NGC 1741-all; h – NGC 3049-1; i – NGC 3125-1; j – NGC 3310-1; k – NGC 3310-all; l – NGC 4214-1; m – NGC 4449-all; n – NGC 4670-1; o – NGC 5102-1; p – NGC 5253-5; q – NGC 5996-1; r – NGC 6764-1; s – NGC 7552-1; t – Tol 1924-416-1; u – Tol 89-1; v – R136. "-1" or "-5" refer to the brightest cluster in the galaxies; "-all" is the average over all clusters covered by the aperture. Sources: Crowther et al. (2016) for R136; all others are from Chandar et al. (2004).

# Tables

Table 1. Adopted properties of II Zw 40.

| Property | $E(B-V)_{MW}$[a] (mag) | $v_{rad}$[b] (km s$^{-1}$) | $D$[c] (Mpc) | $M_B$[d] (mag) | $M_{dyn}$[e] ($M_\odot$) | log(O/H)+12[f] |
|---|---|---|---|---|---|---|
| Value | 0.73 | 789 | 11.1 | −18.1 | $6 \times 10^9$ | 8.09 |

a – Galactic foreground reddening (Schlafly & Finkbeiner 2011)
b – heliocentric radial velocity (NED)
c – distance from $v_{rad}$ and the velocity field model of Mould et al. (2000)
d – absolute B magnitude from Gil de Paz et al. (2003)
e – dynamical mass from H I observations (Brinks & Klein 1988)
f – oxygen abundance from Guseva et al. (2000)

Table 2. Derived properties of SSC-N.

| Property | $M$[a] ($M_\odot$) | $T$[b] (Myr) | $E(B-V)_{int}$[c] (mag) | $L_{Bol}$[d] ($L_\odot$) | $N_L$[e] (s$^{-1}$) | SN rate[f] (yr$^{-1}$) | N(W-R)[g] | log $U_0$[h] | log C/O[i] |
|---|---|---|---|---|---|---|---|---|---|
| Value | $9 \times 10^5$ | 2.8 | 0.07 | $1 \times 10^9$ | $6 \times 10^{52}$ | $<3 \times 10^{-4}$ | 1600 | −2 | −0.70 |

a – stellar mass from the UV luminosity
b – age from the UV line fitting
c – intrinsic reddening from the UV slope
d – bolometric luminosity from the $M/L$ ratio
e – number of ionizing photons predicted from the fit to the UV spectrum
f – supernova rate predicted for a cluster of the derived mass and age
g – number of W-R stars if He 1640 is indicative of WN stars with $L_{1640} = 7 \times 10^{36}$ erg s$^{-1}$
h – ionization parameter from the UV emission lines
i – carbon over oxygen ratio